\documentstyle[amssymb,prl,aps,epsfig,twocolumn]{revtex}

\draft

\begin{document}

\twocolumn[\hsize\textwidth\columnwidth\hsize\csname
@twocolumnfalse\endcsname

\title{Chaotic Motion  Around  Prolate 
Deformed Bodies}

 \author{Eduardo Gu\'eron\thanks{e-mail: gueron@ime.unicamp.br} 
 and
   Patricio S. Letelier\thanks{e-mail: letelier@ime.unicamp.br} }

\address{
Departamento de Matem\'atica Aplicada, Instituto de Matem\'atica,
Estat\'{\i}stica e Computa\c{c}\~ao Cient\'{\i}fica, Universidade 
Estadual de
Campinas, 13083-970, Campinas, SP, Brazil
}
 \maketitle
 
\begin{abstract}
 The motion  of  particles  in the
 field of forces associated to an axially symmetric attraction
 center modeled by   a monopolar term  plus a prolate quadrupole 
 deformation are studied using 
 Poincar\'e surface of sections  and Lyapunov characteristic
numbers.  We find chaotic motion  for certain
  values of the parameters, and that the instability of the orbits
 increases when  the quadrupole parameter increases.
  A general relativistic analogue is  briefly discussed.
\end{abstract}
 
] 

Attraction forces represented by a monopolar plus a prolate quadrupolar
distribution of masses (charges) are a good approximation for elongated
massive (charged) bodies. Examples range from astrophysics to nuclear
physics. There are many observed galaxy clusters with a cigar like shape 
\cite{cooray}. Also, the nuclear charge of light gold atoms has been
reported as having a large prolate deformation \cite{au}. Most of the Dwarf
Galaxies in the Virgo Cluster may obey the ``prolate hypothesis'', i.e.,
they probably have a prolate spheroidal shape\cite{ryden}.
Asteroids also have a prolate shape, but usually they are not axisymmetric.  
Merrit\cite{merr} found,  from detailed  modeling of triaxial galaxies, that
 most of the galaxies must be nearly axisymmetric, either prolate or oblate.

Classical, as well as, quantum chaos have been studied in a variety of
axially symmetric fields of forces. In particular, attraction centers
described by potentials that are the sum of two terms: a monopolar term and
a quadrupolar deformation. Furthermore this center is ``perturbed'' by an
external distribution of masses (charges) represented by its external
multipolar moments, i.e., 
\begin{eqnarray}
&&V=-\alpha /R-qP_{2}(\cos \vartheta )/R^{2}+V_{P}, \\
&&V_{P}=Q_{1}RP_{1}(\cos \vartheta )+Q_{2}R^{2}P_{2}(\cos \vartheta )+...\;.
\end{eqnarray}
Sometimes the monopolar term is changed by the potential of a spring \cite
{heiss}. In general, in all these cases the terms that originate the chaos
are the external multipolar moments.

We shall consider the simplest, albeit, important case of a particle moving
in the field of a monopole plus a quadrupole deformation. This deformation
is usually considered to be the major deviation from spherical symmetry. In
cylindrical coordinates, $(r,\varphi ,z)$, the field takes the generic form, 
\begin{equation}
U(r,z)=-\frac{\alpha }{\sqrt{r^{2}+z^{2}}}-\frac{q(2z^{2}-r^{2})}{%
2(r^{2}+z^{2})^{5/2}},  \label{pot}
\end{equation}
where $\alpha $ is a constant that may be associated with the central body
mass (charge). It  is instructive to have a special model in mind, consider
two equal masses located on  the $z$-axis symmetrically, at $z=-a$ and $z=+a$.
The gravitational potential of the above mass configuration up to the order
 $a^3$ is (\ref{pot}) with $q=2\alpha a^2$. 
We shall use $\alpha =1$ without loss of generality. Note
that we are not considering external multipolar moments ($V_{P}=0$),
i.e., only deformed cores will be studied. 

We can distinguish two cases depending on the sign of $q$. The oblate
deformation case, $q<0$. This is the common case for bodies deformed by
rotation and has been analyzed in astronomy for more than two hundred years. 
The integrability of the Newton equations for a particle moving in
the gravitational field of an axially symmetric oblate body is an unsolved
problem. It is known as the classical problem of the existence of the third
isolating integral of motion \cite{boca}. There are numerical evidences that
orbits of particles moving around a monopole plus an oblate quadrupole are
not chaotic.  We study the prolate deformation
case, $q>0$; that is the one that we shall discuss in this communication. In
this case we have a monopolar field (the usual Kepler problem) ``perturbed''
by a quadrupolar term, in other words, we have a typical situation wherein
the KAM (Kolmogorov-Arnold-Moser) theory applies \cite{gh}.

First we study the contours of the effective potential $%
U_{eff}=U+h_{z}^{2}/(2r^{2}),$ where $h_{z}=r^{2}\dot{\varphi}$ is the axial
specific angular momentum that due to the axial symmetry is conserved. We
also have the conservation of the total specific energy, $E=(\dot{r}^{2}+%
\dot{z}^{2})/2+U_{eff}$. Thus we have that the motion is completely
determined by the functions $r=r(t)$ and $z=z(t)$. Then, we have a four
dimensional phase space. But, due to energy conservation the motion actually
takes place in a three dimensional space. An adequate tool to investigate
the trajectories in this phase space is the Poincar\'{e} surface of section
method. Now let us comeback to the effective potential contours. In Fig. 1
we plot the level contours of $U_{eff}$ for $L_{z}=0.83$ and $E=0.464$ and
different values of the quadrupole moment parameter: a) $q=0.3$, b) $q=0.5$,
c) $q=0.85$, and d) q=0.95. Thus for these values of the parameters the
motion of the particle is confined to toroidal regions that do not contain
the symmetry axis. Note that for the last case we have two non connected
regions.

The particles move in the reduced phase space $(p_{r}=\dot{r},r,z)$. Note
that $p_{z}=\dot{z}$ is determined by the energy conservation. In Fig. 2,
for the case a), we present the intersection points of some particle
trajectories with the plane $z=0$. The picture is the one for regular
orbits. The case b) is analyzed in Fig. 3, using the same surface section.
We find regions of non destroyed tori together with chaotic regions in
concordance with the KAM theory. In Fig. 4 we show again the case b) but,
now with a different section, $z=0.4$. We see that the integrable and
chaotic regions are deformed depending on the chosen section. We also
studied the case c) that is quite similar to the former, so we shall not
present it here. We find that increasing the quadrupole moment the size of
the chaotic regions also increases. And finally, in Fig. 5 we study orbits
in one of the non connected regions of the case d). In this last case the
surface section is taken as $z=0.4$, again we find large regions of chaotic
behavior and some non destroyed tori. In summary, we find chaotic behavior
of orbits for several values of prolate quadrupole moment.

 To quantify the degree of instability of the orbits  we  shall study
their associated 
 Lyapunov characteristic numbers (LCN) that are defined as the double 
limit 
\begin{equation}
LCN=\lim_ 
{\scriptsize\begin{array}{l}\delta _{0}\rightarrow 0  \\ t\rightarrow \infty\end{array}} 
 \left[ 
{{\log (\delta /\delta _{0}) \over t}}%
\right] , 
\end{equation}
where $\delta _{0}${ \ and }$\delta ${\ are the deviation of
two nearby orbits at times }$0${ \ and }$t${ \ respectively. We get
the largest LCN by using the technique suggested by Benettin et al. \cite
{bgs}}

 We fix the value of the  constants
of motion as  $L_{z}=0.83$ and $E=0.464$,  and choose the same values of
quadrupole parameters  used to plot the Poincar\'{e} sections.
 For the value  $q=0.3$ it
was chosen the reference orbit with initial conditions:
 $z=0$, $p_{r}=0$, and $r=0.85$,  and $%
\delta _{0}\simeq 10^{-9}$,  we found $LCN\lesssim 10^{-4}$ \
that characterizes  a stable system. With $q=0.5$ and
 initial conditions:  $z=0$, $
p_{r}=0.2,$
\ Finally, for $q=0.95$, and  $z=0.4$ , $p_{r}=0.05,$ and $r=0.95$, \ we
obtain $LCN\simeq 0.09$\ $(\pm 0.015).$ \ We see  that the
degree of instability  increases when  the quadrupole parameter
increases for fixed constants of motion.

As we said before, there are numerical evidences that
orbits of particles moving around a monopole plus an
 oblate quadrupole are
not chaotic.  The difference between the oblate and the prolate case
can be understood  analyzing the critical points of 
the effective potential $U_{eff}$. In particular,
the  existence of the saddle points that is one of the 
main ingredients to have instable motion. We find 
  that the   critical point, $r=\sqrt{(3q+2L_{z}^{2})/2\alpha }$, $z=0, \;$
  is a saddle point if the parameters obey the two conditions, $L_{z}^{2}<3q$ \ and $3L_{z}^{2}>%
\sqrt{2\alpha /3}.$ \ therefore when $q<0,$ \ the oblate case, no
real $L_{z}$ \ can obey the first of these condition. 

The Newtonian motion of a particle moving in the potential (3) has
 a general
relativistic analogue. The potential is replaced by a metric
solution to the vacuum Einstein equation  and the particle
 motion equation by the geodesic equation. The instability of geodesics in metrics associated to
a black hole surrounded by a shell of matter was studied  in some
detail in \cite{vl}.

 A solution to the Einstein equations that has as a Newtonian limit a potential like (\ref{pot}) is the Erez-Rosen-Quevedo (ERQ) 
solution \cite{quevedo}. We
did not found chaos in the oblate case, the prolate case is also chaotic. 
 The confinement region for  the relativistic motion constants,  $E=0.937$,
 and \
 $L_{z}=3.322$, \ and the quadrupole parameter $q=5.02$ is presented in Fig. 6.
The coordinates used in this case are spheroidal coordinates,
that are the ones appropriate for the ERQ solution, they are related to the
usual cylindrical coordinates by $u= (R_{+} + R_{-})/(2m),$ and $ v=(R_{+} - R_{-})/(2m)$, with $R_\pm=(r^2+(z\pm m)^2)^{1/2}.$ 
We have two regions of confinements that we have labeled II and III. In Fig. 7
we present a Poincar\'e section for particles moving in the region II, 
the section is taken as $v=0$, i.e., $z=0$;  $u, p_u$ are  canonical
 conjugate variables. We see a phase space with 
a large region of chaotic motion.  We shall present a
 complete study of geodesic in ERQ spacetimes elsewhere.

Some dense cores in dark clouds have been found to have prolate spheroidal
shape\cite{myers}. Then a prolate geometry has to be considered as initial
condition in the star formation process. We think that the strong
instability presented here  may play a crucial effect in the formation
of structures in starts \cite{w}.

We want to finish this short communication by reminding that in nonlinear
systems of equations chaos is the rule rather than the exception. Thus
simple systems with a minimum of structure play an important role in the
physical, as well as, mathematical understanding of chaos. A good example is
the paradigmatic Hen\'{o}n-Heiles system wherein the ``simple'' addition of
a $x^{2}y$ term in the potential of two uncoupled oscillators (integrable
motion) has dramatic consequences that are the physical manifestation of the
creation of a saddle point together with a perturbation. In the case
presented here, we have a similar situation, the prolate-quadrupole
potential also adds a saddle and a perturbation.

The authors thank CNPq and FAPESP for financial support and M.A.M. Aguiar
(IFGW-UNICAMP) for several discussions concerning chaos.


\begin{figure}[tbp]
\epsfig{width=3in,height=2in, file=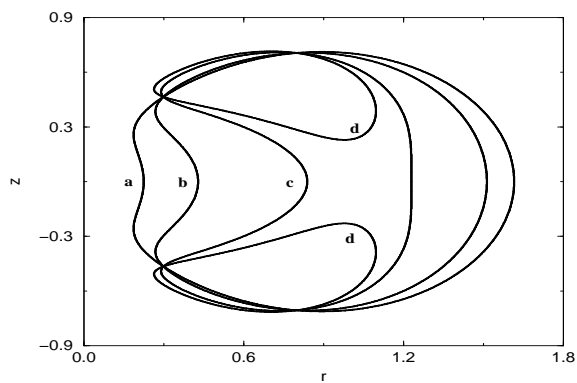} 
\caption{We plot the level contours of $U_{eff}$ for $L_{z}=0.83$ and $%
E=0.464$ and different values of the quadrupole moment parameter: a) $q=0.3$%
, b) $q=0.5$, c) $q=0.85$, and d) q=0.95.}
\end{figure}

\begin{figure}[tbp]
\epsfig{width=3in,height=2in, file=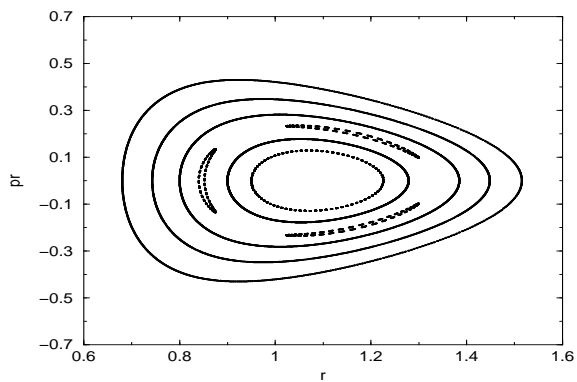} 
\caption{Surface of section for $L_{z}=0.83$ and $E=0.464$ and $q=0.3$. The
section corresponds to the plane $z=0$. For these values of the parameters
we have the section of regular motion.}
\end{figure}
\begin{figure}[tbp]
\epsfig{width=3in,height=2in, file=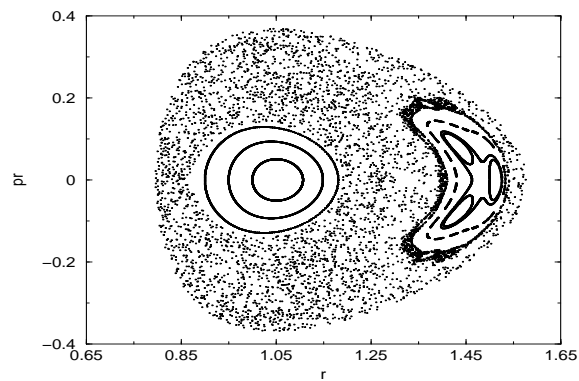} 
\caption{Surface of section for $L_{z}=0.83$ and $E=0.464$ and $q=0.5$. The
section corresponds to the plane $z=0$. For these values of the parameters
we have the typical section indicating chaotic motion.}
\end{figure}

\begin{figure}[tbp]
\epsfig{width=3in,height=2in, file=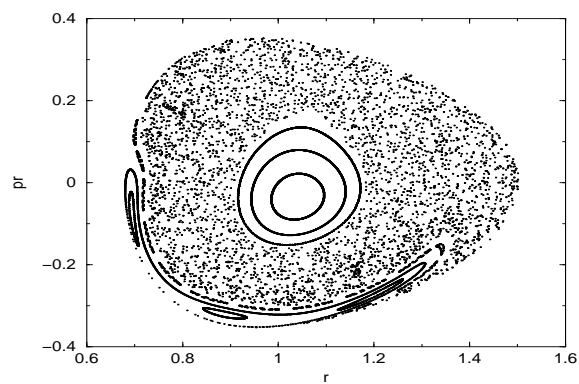} 
\caption{ Surface of section for the same values of the parameters that in
the precedent figure, but a different section, $z=0.4$. We see a different
cut of the regular and chaotic regions. }
\end{figure}

\begin{figure}[tbp]
\epsfig{width=3in,height=2in, file=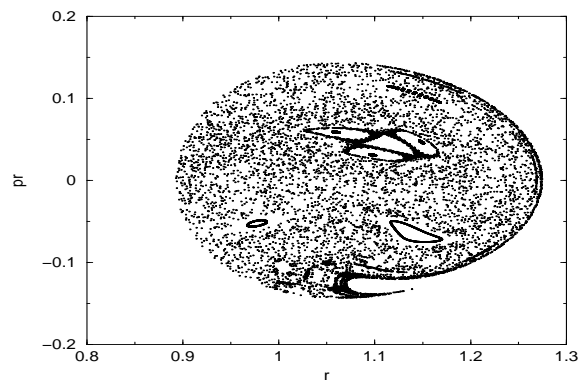}  
\caption{ Surface of section for $L_{z}=0.83$ and $E=0.464$ and $q=0.95$.
The section corresponds to the plane $z=0.4$. Again we have irregular
motion. }
\end{figure}
\begin{figure}[tbp]
\epsfig{width=3in,height=2in, file=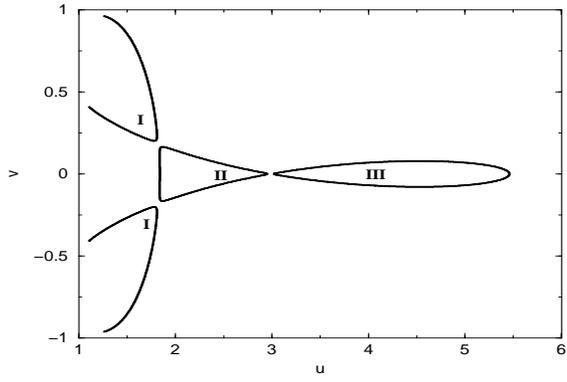}  
\caption{Level contour for the general relativistic quadrupole + monopole
  system (ERQ solution). The relativistic constants are $L_{z}=3.32$ and 
$E=0.937$ and $q=5.02$. The labels $u$ and $v$ denote spheroidal coordinates.
  }
\end{figure}
\begin{figure}[tbp]
\epsfig{width=3in,height=2in, file=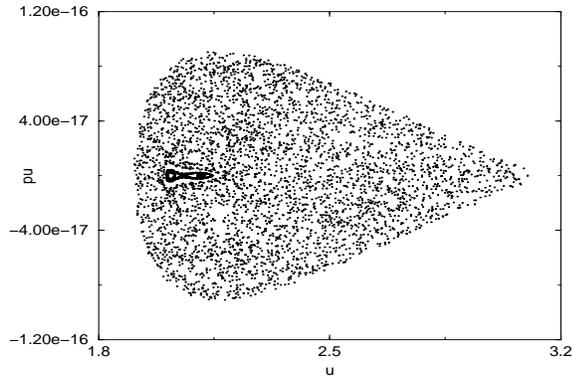}  
\caption{ Surface of section for the region II shown in previous figure. 
We have a large region of  chaotic motion. The section
 corresponds to the plane $v=0$ i.e., $z=0$.}
\end{figure}

\begin{references}
\bibitem{w}  Work along this line will be soon reported.
\end{references}
\end{document}